\documentclass[reprint,superscriptaddress,nofootinbib,amsmath,amssymb,aps,prl]{revtex4-2}

\usepackage{float}
\usepackage{graphicx}
\usepackage{amssymb}
\usepackage{amsmath}
\usepackage{hyperref}
\usepackage{dcolumn}   
\usepackage{bm}        
\usepackage{color}
\usepackage{appendix}
\usepackage[bottom]{footmisc}
\newcommand{\be}{\begin{equation}}
\newcommand{\en}{\end{equation}}
 \newcommand{\bea}{\begin{eqnarray}}
 \newcommand{\ena}{\end{eqnarray}}

\begin{document}

\title{Quantum anomaly triggers the violation of 
scaling laws in gravitational system}
\author{Ya-Peng Hu}
\thanks{These authors contributed equally to the work}
\thanks{Correspondence author}
\email{huyp@nuaa.edu.cn}
\affiliation{
 College of Physics, Nanjing University of Aeronautics and Astronautics, Nanjing, 211106, China
  }
\affiliation{Key Laboratory of Aerospace Information Materials and Physics (NUAA), MIIT, Nanjing 211106, China}
\author{Yu-Sen An}
\thanks{These authors contributed equally to the work.}
\email{anyusen@nuaa.edu.cn}
\affiliation{
 College of Physics, Nanjing University of Aeronautics and Astronautics, Nanjing, 211106, China  }
\affiliation{Key Laboratory of Aerospace Information Materials and Physics (NUAA), MIIT, Nanjing 211106, China}
\author{Gao-Yong Sun}
\affiliation{
 College of Physics, Nanjing University of Aeronautics and Astronautics, Nanjing, 211106, China  }
\affiliation{Key Laboratory of Aerospace Information Materials and Physics (NUAA), MIIT, Nanjing 211106, China}
\author{Wen-Long You}
\affiliation{
 College of Physics, Nanjing University of Aeronautics and Astronautics, Nanjing, 211106, China  }
\affiliation{Key Laboratory of Aerospace Information Materials and Physics (NUAA), MIIT, Nanjing 211106, China}
\author{Da-Ning Shi}
\affiliation{ College of Physics, Nanjing University of Aeronautics and Astronautics, Nanjing, 211106, China }
\affiliation{Key Laboratory of Aerospace Information Materials and Physics (NUAA), MIIT, Nanjing 211106, China}
\author{Hongsheng Zhang}
\thanks{Correspondence author}
\email{sps\_zhanghs@ujn.edu.cn}
\affiliation{School of Physics and Technology, University of Jinan, 336 West Road of Nan Xinzhuang, Jinan, Shandong 250022, China }
\author{Xiaosong Chen}
\thanks{Correspondence author}
\email{chenxs@bnu.edu.cn}
\affiliation{ School of Systems Science, Beijing Normal University, Beijing 100875, China }
\author{Rong-Gen Cai}
\thanks{Correspondence author}
\email{cairg@itp.ac.cn}
\affiliation{School of Physics Sciences and Technology, Ningbo University, Ningbo, 315211, China}
\affiliation{School of Fundamental Physics and Mathematical Sciences, Hangzhou Institute for Advanced Study, UCAS, Hangzhou 310024, China}
\affiliation{CAS Key Laboratory of Theoretical Physics, Institute of Theoretical Physics, Chinese Academy of Sciences, P.O. Box 2735, Beijing 100190, China}

\begin{abstract}
Scaling laws for critical phenomena take pivotal status in almost all branches of physics. However, as scaling laws are commonly guaranteed by the renormalization group theory, systems  that violate them have rarely been found. In this letter, we demonstrate that gravitational system can break scaling laws. We derive this result through investigating phase transition and critical phenomenon in a gravitational system with quantum anomaly. For the first time, we outline the key conditions to violate the scaling laws in generic gravitational system viewed from the equation of state $P=P(T,V)$. Our results indicate that quantum effects can magnify the distinctiveness of gravity, which may be significant to understand the microscopic structure of spacetime. 
\end{abstract}
\maketitle
{\bf Introduction}---
 Phase transition and critical behaviour are fairly common in nature. It is remarkable that for critical phenomena, universal relations appear among the critical exponents  in diverse systems with different fundamental physical laws, such as van de Waals liquid-gas system and Ising model. These universal relations are now widely known as scaling laws~\cite{Pelissetto:2000ek}. 
 Later it was further found that this universality of scaling laws roots in the scale-invariant
 property of the critical point with long-range correlation, which is independent of the microscopic details of the system. Moreover, this universality can also be understood in terms of the renormalization group
 theory 
 \cite{Wilson:1974mb}.

 It is a remarkable result that black hole, a fairly simple object in gravity theory, can also exhibit phase transition and critical phenomenon which relate to the discovery of black hole thermodynamics. 
 Original black hole thermodynamics was established in Einstein gravity for asymptotic flat spacetime after the discoveries of black hole mechanics\cite{Bardeen:1973gs}, Hawking radiation \cite{Hawking:1975vcx} and Bekenstein-Hawking entropy\cite{Bekenstein:1973ur}. 
 For asymptotic AdS spacetime with negative cosmological constant, black hole thermodynamics has presented various distinct phenomena, such as the famous Hawking-Page phase transition \cite{Hawking:1982dh} which corresponds to confinement-deconfinement phase transition in gauge theory viewed from AdS/CFT correspondence \cite{Witten:1998zw}. Therefore, much attention has been paid on the AdS black hole thermodynamics and its duality interpretation \cite{Chamblin:1999tk,Chamblin:1999hg,Caldarelli:1999xj,Zhao:2023gur,Li:2022oup}. Interestingly, if one further identifies the negative cosmological constant to be thermodynamic pressure of the AdS black hole spacetime \cite{Dolan:2011xt} and incorporates the variation of cosmological constant, AdS black holes have manifested more abundant thermodynamic behaviors. For example, van der Waals like phase transition \cite{Kubiznak:2012wp}, reentrant phase transition \cite{Altamirano:2013ane}, black hole triple point \cite{Altamirano:2013uqa}, super-fluid black hole phase \cite{Hennigar:2016xwd} and multi-critical points \cite{Tavakoli:2022kmo}. It should be pointed out that in this case the critical phenomenon with continuous phase transition can also occur in AdS black hole thermodynamics, where the critical exponents are the same as the result of mean field theory and adhere to scaling laws for almost all cases \cite{Kubiznak:2016qmn}.\footnote{However, see a possible violation in Refs.\cite{Tavakoli:2022kmo,Frassino:2014pha,Dolan:2014vba}. Note that these phenomena occur in seven dimensional hyperbolic Lovelock black hole,  whether it makes realistic sense is subtle. 
 }
Note that, the scaling laws can be derived by renormalization group method in the framework of local quantum field theory, while gravitational theory at high energy scale can contain non-local effect which may not be consistent with the framework of local quantum field theory \cite{Solodukhin:1997yy,Aros:2010jb}.Therefore, a very interesting and important question is to ask whether the critical exponents of systems controlled by gravity always satisfy the scaling laws. 

In this Letter, we demonstrate that gravitational system with quantum anomaly has the ability to break the scaling laws, and hence make a significant step towards the answer to the above question.
Different from the black hole thermodynamics 
in the context of classical gravity, gravitational system with quantum effects has already shown some unique properties. 
For example, entropy can be significantly modified with logarithmic correction \cite{Kaul:2000kf,Solodukhin:1997yy,Carlip:2000nv,Das:2001ic}, and the  Hawking radiation can be derived by gravitational anomaly cancellation to avoid the breakdown of general covariance at quantum level \cite{Robinson:2005pd}. 
Besides the above effects on thermodynamic properties, gravitational system itself can also be modified with quantum correction. Particularly, static spherically symmetric black hole solutions with quantum anomaly have been found in Ref.\cite{Cai:2009ua,Cai:2014jea}, for generalization to rotating and non-stationary case see Ref.\cite{Fernandes:2023vux,Gurses:2023ahu}. Based on these black hole solutions,  
we have further discovered for the first time the violation of scaling laws in four-dimensional gravitational system by investigating critical phenomena for black hole with quantum anomaly. In addition, we find a novel phase transition behavior, namely that first-order phase transitions occur both above and below the critical temperature. Furthermore, we first obtain the key condition to violate the scaling laws for a generic gravitational system with the equation of state $P=P(T,V)$. 
This condition is presented as a relationship between the U (1) charge $Q$ and type A central charge $\alpha_{c}$ in the black hole with quantum anomaly case. Our results reveal that quantum anomaly can magnify the distinctiveness of gravity and trigger the violation of scaling laws, which are insightful to understand the underlying relationship between microscopic structure of spacetime and violation of scaling laws. 



{\bf Black hole solution with quantum anomaly }---
There are several ways to incorporate quantum effects in black hole \cite{Cai:2009ua,Cai:2014jea,Lu:2009em,Ashtekar:1997yu,Garfinkle:1990qj}. One interesting way is to consider the back-reaction caused by conformal anomaly of quantum field theory in black hole spacetime \cite{Cai:2009ua,Cai:2014jea}. For classical field theory with conformal symmetry, the trace of stress energy tensor vanishes.  Nevertheless, for quantum conformal field theory (CFT) in four dimension, its vacuum expectation value is \cite{Duff:1993wm,Deser:1993yx}
\begin{equation}
\label{tanomaly}
    \langle T^{\mu}_{\mu} \rangle_{a}=b I_{4}-a E_{4},
\end{equation}
where the first term reads $I_{4}=C_{\mu\nu\lambda\delta}C^{\mu\nu\lambda\delta}$ with $C_{\mu\nu\lambda\delta}$ representing Weyl tensor, and the second term is Euler characteristic $E_{4}=R_{\mu\nu\lambda \delta}R^{\mu\nu\lambda\delta}-4R_{\mu\nu}R^{\mu\nu}+R^{2}$. $b$, $a$ are two non-negative constants proportional to the central charge which reflect the degrees of freedom of underlying quantum field theory \cite{Duff:1993wm,Deser:1993yx}.
From Eq.(\ref{tanomaly}), it is obvious that the trace of stress tensor is non-zero when the CFT is placed on curved spacetimes, while it is zero in the flat case. The non-zero $\langle T^{\mu}_{\mu}\rangle_{a}$ is called quantum conformal anomaly, where the first term on the RHS of Eq.(\ref{tanomaly}) is referred as type B anomaly and the second term type A anomaly. 

Taking into account the back-reaction caused by this quantum anomaly, the classical Einstein equation will be modified to be
\begin{equation}
    R_{\mu\nu}-\frac{1}{2}g_{\mu\nu}R+\Lambda g_{\mu\nu}=8\pi \langle T_{\mu\nu}\rangle_{a},
\end{equation}
where we use geometric unit $G=c=1$ for convenience. After assuming spherical symmetry and a specific equation of state $\langle T^{t}_{t}\rangle_{a}=\langle T^{r}_{r}\rangle_{a}$, a spherical symmetric black hole solution in the presence of quantum anomaly has been found \cite{Cai:2009ua,Cai:2014jea} as \footnote{Note that this metric is the same as the black hole solution obtained in 4d Einstein-Gauss-Bonnet gravity in Ref.\cite{Glavan:2019inb,Lu:2020iav,Wei:2020poh} up to a simple parameter replacement $\alpha_{c} \to -\alpha/2$, with $\alpha$ representing the Gauss-Bonnet coupling constant. However, the crucial difference is that positive central charge $\alpha_{c}$ in our metric corresponds to negative Gauss-Bonnet coupling $\alpha$ which is excluded by the unitary constraint \cite{Cheung:2016wjt}. Several researches have demonstrated the dynamical stability of this solution with positive $\alpha_{c}$ \cite{Zhang:2020sjh,Aragon:2020qdc,Konoplya:2020bxa}. }
\begin{equation}
    ds^{2}=-f(r)dt^{2}+\frac{1}{f(r)}dr^{2}+r^{2}(d\theta^{2}+\sin^{2}\theta d\phi^{2}).
\end{equation}
and
\begin{equation}
    f(r)=1-\frac{r^{2}}{4\alpha_{c}}(1-\sqrt{1-8\alpha_{c}(\frac{2M}{r^{3}}-\frac{Q^{2}}{r^{4}}-\frac{1}{l^{2}})})\label{solu}
\end{equation}
where only type A anomaly is considered with $\alpha_{c}=8\pi a$ and $b$ set to zero. $M$ and $Q$ are two integration constants, and $l$ is related to cosmological constant by $\Lambda=-\frac{3}{l^{2}}$.
The trace anomaly equation (\ref{tanomaly}) and the assumption $\langle T^{t}_{t}\rangle_{a}=\langle T^{r}_{r}\rangle_{a}$ are crucial to obtain this spherical black hole solution whose physical meaning can be illuminated from the effective action perspective. The effective action which captures conformal anomaly has been proposed by Ref.\cite{Riegert:1984kt,Komargodski:2011vj} which takes the Horndeski form. By varying the effective action, effective stress energy tensor will be found and the particular form of Horndeski scalar profile can be computed which makes the trace anomaly equation and condition $\langle T^{t}_{t}\rangle_{a}=\langle T^{r}_{r}\rangle_{a}$ satisfied \cite{Fernandes:2021ysi}. We have also checked the solution satisfies all the other components of the semi-classical Einstein equations. Therefore, the metric in Eq.(\ref{solu}) represents a truly physical special solution with conformal anomaly.

When the quantum correction is small, the above black hole solution reduces to RN-AdS black hole $f(r)=1-\frac{2M}{r}+\frac{Q^{2}}{r^{2}}+\frac{r^{2}}{l^{2}}$, thus it has been demonstrated that parameter $M$ is the mass of the black hole, $Q$ can be interpreted as the $U(1)$ charge of underlying conformal field theory \cite{Cai:2009ua,Cai:2014jea}.
The mass and temperature of the black hole can be written in terms of outer horizon radius $r_{h}$ and charge $Q$ as \cite{Cai:2009ua,Cai:2014jea,Wei:2020poh}
\begin{eqnarray}
M&=&\frac{ r_{h}^{4}/l^2+r_{h}^{2}+Q^{2}-2\alpha_{c}}{2r_{h}},\nonumber\\
T&=&\frac{3 r_{h}^{4}/l^2+r_{h}^{2}-Q^{2}+2\alpha_{c}}{4\pi r_{h}^{3}-16\pi \alpha_{c} r_{h}}.
\end{eqnarray}
Meanwhile, the black hole also satisfies following thermodynamic first law
$dM=TdS+\Phi dQ$, 
where the entropy $S$ and chemical potential $\Phi$ are respectively
\begin{equation}
    S=\int \frac{dM}{T}=\frac{A}{4}-4\pi \alpha_{c} \ln(\frac{A}{A_{0}}),\quad \Phi=Q/r_{h},
\end{equation}
and $A$ is the black hole horizon area $4\pi r_{h}^{2}$, while $A_{0}$ is an integration constant. 

From the above results, 
it is obvious that the quantum anomaly indeed affects the thermodynamic variables, including mass, temperature and entropy.
Interestingly, the well known logarithmic correction of black hole entropy due to quantum effect also appears \cite{Kaul:2000kf}.
Note that one can not uniquely fix the $A_{0}$, which depends on the microscopic details of quantum gravity theory.
Usually, one sets $A_{0}=8\pi \alpha_{c}$ for simplicity, so we also use this setting and the reader can consult more discussions about $A_{0}$ in Ref.\cite{Cai:2014jea}.

{\bf Quantum effect on black hole critical phenomena}---\label{thermal}
For AdS black hole systems, after further identifying cosmological constant as the thermodynamic pressure $P=-\frac{\Lambda}{8\pi}=\frac{3}{8\pi l^{2}}$,
one can obtain more abundant thermodynamic phenomena such as  critical phenomena~\cite{Kubiznak:2012wp}. 
Therefore, discovery of  quantum effects on the black hole critical phenomena would be important.


In our case for AdS black hole with quantum anomaly, 
the extended thermodynamic relation is
\begin{equation}
    dM=TdS+\Phi dQ +VdP, \label{First Law}
\end{equation}
where $V$ is the conjugate thermodynamic volume $V=4\pi r_{h}^{3}/3$, and the mass $M$ is interpreted as enthalpy $H$ instead of internal energy, i.e. $H=M$~\cite{Kubiznak:2016qmn,Kastor:2009wy}. Moreover, one can easily obtain the equation of state in the extended phase space 
\begin{eqnarray}
	P=\frac{T}{2 r_h}-\frac{1}{8 \pi  r_h^2}-\frac{2\alpha_{c}T}{r_h^3}+\frac{-2\alpha_{c} +Q^2}{8 \pi  r_h^4},\label{P}
\end{eqnarray}
which is same as the case in the four-dimensional charged Gauss-Bonnet black hole in AdS space~\cite{Wei:2020poh}. However, the physical meaning of $\alpha_{c}$ and $Q$ are different.  

We focus on the canonical ensemble with charge $Q$ fixed throughout this letter. Therefore, Eq.(\ref{First Law}) reduces to the standard first law of thermodynamics $dH=TdS+VdP$. 
After using the conditions of critical point 
$\left[\left(\dfrac{\partial P}{\partial V}\right)_T\right]_c=\left[\left(\dfrac{\partial ^2P}{\partial
V^2}\right)_T\right]_c=0$, the critical values are calculated as
~\cite{Wei:2020poh} 
\begin{eqnarray}\label{crit}
V_c&=&\frac{4\pi}{3}(-12\alpha_{c}+3Q^2 + K)^{3/2},\nonumber\\	T_c&=&\frac{3Q^2 - K}{48\pi \alpha_{c} \sqrt{-12\alpha_{c} +3Q^2+ K}}, \nonumber\\ 
P_c&=&\frac{-18\alpha_{c} +6Q^2 + K}{24\pi(-12\alpha_{c}+3Q^2+ K)^2},	
\end{eqnarray}
where $K\equiv\sqrt{192\alpha_{c}^2+9Q^4-96\alpha_{c} Q^2}$ and $V_c=4\pi r_{hc}^3/3$. 
It should be emphasized that the above critical values are only satisfied under the condition $\alpha_{c} \leqslant 0$. While if one considers the non-negative value of central charge and existence of critical point, the region of physical parameter should be $0 \leqslant \alpha_{c} \leqslant \frac{Q^{2}}{8}$. In this physical case, in addition to the aforementioned critical point, there is another critical point where $V_c=\frac{4\pi}{3}(-12\alpha_{c}+3Q^2 - K)^{3/2}$, $T_c=\frac{3Q^2 + K}{48\pi \alpha_{c} \sqrt{-12\alpha_{c} +3Q^2- K}}$ and $P_c=\frac{-18\alpha_{c} +6Q^2 - K}{24\pi(-12\alpha_{c}+3Q^2- K)^2}$. 
However, when $\alpha_{c}$ approaches zero, this another critical point exhibits divergent behaviour, while the above critical values in (\ref{crit}) can return to the results in RN-AdS black hole, thus we will only focus our investigations around the critical point in (\ref{crit}).



In the following, we aim to make our discussions of critical phenomena as general as possible, where only two conditions are required. First, the standard first law of thermodynamics holds, i.e. $dH=TdS+VdP$. Second, at least one critical point exists in the $P$-$V$ phase transition.
Therefore, we can expand the thermodynamic pressure $P(T,V)$ around the critical point
	\begin{eqnarray}\label{pressure}
		\tilde{P}(t,w)&=&1+\sum_{i\geq 1}a_{i0}t^{i}+\sum_{j\geq 3}a_{0j}w^{j}+\sum_{i,j\geq 1}a_{ij}t^{i}w^{j}, \quad \label{Pexapnsion}
	\end{eqnarray}
	where $\tilde{P}\equiv P/P_c$, $t\equiv T/T_c-1$, $w \equiv V/V_c-1$, 
 and $a_{ij}$   representing the coefficients of $t^i w^j$ terms. Note that $a_{01}$ and $a_{02}$ vanish due to the existence of critical point. In addition, the stability condition of thermodynamic systems keeping  Gibbs free energy having minimum requires $(\partial^{3}P/\partial V^{3})_{T=T_{c}}<0$ as discussed in Ref.~\cite{Landau:1980mil}, and hence $a_{03}=\frac{V_{c}^{3}}{P_{c}^{3}}(\partial^{3}P/\partial V^{3})_{T=T_{c}}<0$ needs to be satisfied throughout the letter.

For the $P$-$V$ phase transition with second order, four critical exponents are defined at the critical point as \cite{Pelissetto:2000ek}
\begin{alignat}{1}
&C_{V}=T(\frac{\partial S}{\partial T})_{V}\propto |t|^{-\alpha},
\\
&\eta=\frac{V_l-V_s}{V_c}\propto |t|^{\beta},
\\
&\kappa_T=-\frac{1}{V}(\frac{\partial V}{\partial P})_T\propto |t|^{-\gamma},
\\
&\tilde{P}-1\propto w^{\delta},
\end{alignat}
where $\alpha$ denotes the divergent behavior of heat capacity at constant volume, $\beta$ describes the behavior of order parameter near the critical temperature, $\gamma$ describes the behavior of isothermal compressibility near the critical point and $\delta$ is related to the shape of isotherm along the critical temperature. 

Taking Maxwell's equal area law into account, which is actually derived from the standard first law $dG=-SdT+VdP$ at fixed temperature, one obtains
\begin{equation}\label{m1}
       \tilde{P}(t,\omega_{l})=\tilde{P}(t,\omega_{s}),~~
       \int_{\omega_{s}}^{\omega_{l}} \omega d\tilde{P}=0
   \end{equation}
where the first equation comes from equal pressure between phase $l$ and $s$, and second equation follows from equal Gibbs free energy during phase transition. The volume $w_l$ and $w_s$ of two phases are calculated from Eq.(\ref{m1}), and hence one can obtain the order parameter such as $\eta=w_{l}-w_{s}$ which are helpful to extract critical exponents.

For the case of a non-zero coefficient $a_{11}$, the two volumes $w_l$ and $w_s$ are easily calculated at leading order as~\cite{Majhi:2016txt,Hu:2020pmr}
\begin{equation}\label{lead}
    w_{s}=-\sqrt{-\frac{a_{11}}{a_{03}}}|t|^{1/2}, \quad w_{l}=\sqrt{-\frac{a_{11}}{a_{03}}}|t|^{1/2}.
\end{equation}
where $a_{11}=\frac{V_{c}T_{c}}{P_{c}^{2}}(\frac{\partial^{2}P}{\partial T \partial V})_{c}$ and $a_{03}$ is negative from stability condition.
Therefore, the critical exponent $\beta$ is read off to be $\beta=1/2$, while the critical exponents $\gamma$ and $\delta$ can also be easily computed as $\gamma=1$, $\delta=3$~\cite{Majhi:2016txt,Hu:2020pmr}. Considering the volume $V$ and entropy $S$ are not independent for spherical symmetric black hole, i.e. $S=A/4$ ~\cite{Bekenstein:1973ur} or $S= A/4+c \log(A)$ with constant $c$ when considering quantum effects~\cite{Kaul:2000kf}. One can prove that $C_{V}=T(\frac{dS}{dT})_{V}$ is always zero, and hence $\alpha=0$. Therefore, the four critical exponents are the same as the mean field theory
\begin{equation}
    \alpha=0,\quad \beta=1/2, \quad \gamma=1,\quad \delta=3. \label{Mean}
\end{equation}
and obey the scaling laws
\begin{eqnarray}
     \alpha+2\beta+\gamma=2,\quad \alpha+\beta(1+\delta)=2,\nonumber\\ \quad \gamma(1+\delta)=(2-\alpha)(\delta-1),\quad \gamma=\beta(\delta-1), \label{scalingl}
\end{eqnarray}
where only two relations are independent. 

For our black hole  with quantum anomaly case, after considering the pressure expression Eq.(\ref{P}), the critical values Eq.(\ref{crit}) and the expansion around critical point Eq.(\ref{pressure}), the leading expansion coefficients are
\begin{eqnarray}
&a_{10}&=\frac{4(-4\alpha_{c}+ K+3Q^{2})}{-22\alpha_{c}+9Q^{2}},~a_{11}=\frac{-48\alpha_{c}- 10 K+6Q^{2}}{-66\alpha_{c}+27Q^{2}}, \nonumber\\ &a_{03}&=\frac{40\alpha_{c}+ K-15Q^{2}}{27(-22\alpha_{c}+9Q^{2})}. \label{coe}
\end{eqnarray}
From Eq.(\ref{coe}), the four critical exponents will be the result of mean field theory in (\ref{Mean}), if $a_{11}$ and $a_{03}$ are not zero as  discussed above. 
Most particularly, we further discover that the $a_{11}$ can be zero if central charge and $U(1)$ charge satisfy $\alpha_{c}=\frac{Q^{2}}{8}$ for this black hole with quantum anomaly case. It should be emphasized that the $a_{11}=0$ case is the first example discovered in four dimensional black hole system. Moreover, this $a_{11}=0$ case can induce notably different critical exponents from mean field theory results, which has not been discussed carefully. Therefore, we will investigate the $a_{11}=0$ case as general as possible like the above discussions in details.

For zero coefficient $a_{11}$ case, Eq.(\ref{m1}) become
\begin{eqnarray}
	&a_{03}( w_l^3-w_s^3)+a_{12}t(  w_l^2-  w_s^2)+a_{21}t^2(  w_l-  w_s)&=0,\nonumber\\
	&\frac{3a_{03}}{4}(w_l^4-w_s^4)+\frac{a_{21}}{2} t^2(w_l^2-w_s^2)+\frac{2 a_{12}}{3} t(w_l^3-w_s^3)&=0.  \quad \quad \label{leading}
\end{eqnarray}
where the original leading order terms related to $a_{11}$ vanish since $a_{11}=0$.
Therefore, the two volumes $w_l$ and $w_s$ are obtained at the leading order as
\begin{equation}
w_l=\text{A}_1 t,\quad  w_s=\text{B}_1 t,\label{w}
\end{equation}
with
\begin{eqnarray}\label{ab}
A_1&=&\frac{\sqrt{3a_{03}^2 \left(a_{12}^2-3 a_{03} a_{21}\right)}-a_{03} a_{12}}{3 a_{03}^2},\nonumber\\
B_1&=&\frac{-\sqrt{3a_{03}^2 \left(a_{12}^2-3 a_{03} a_{21}\right)}-a_{03} a_{12}}{3 a_{03}^2}.
\end{eqnarray}
which are clearly different from the above non-zero $a_{11}$ case in Eq.(\ref{lead}). Near the critical point, we obtain
\begin{eqnarray}
    &\eta& = w_{l}-w_{s}=\frac{V_{l}-V_{s}}{V_{c}}\propto |t|^1, \nonumber\\
    &\kappa &\propto -(\frac{\partial \tilde{P}}{\partial \omega})_{t}^{-1} \propto t^{-2},\nonumber\\
&\tilde{P}&-1\propto \omega^{3},
\end{eqnarray}
which give exponent $\beta=1$, $\gamma=2$ and $\delta =3$. For spherical symmetric black hole systems, the critical exponent $\alpha$ is still zero. 
Thus we obtain the four critical exponents in this second order phase transition case
\begin{eqnarray}\label{ce}
\alpha=0, \quad \beta=1, \quad \gamma=2,\quad \delta=3.
\end{eqnarray}
Comparing with the mean field result in Eq.(\ref{Mean}), we discover that the critical exponents $\beta$ and $\gamma$ are different, while $\alpha$ and $\delta$ are the same.
Most surprisingly, we further discover that the scaling laws for critical phenomena can be violated compared with Eq.(\ref{scalingl}). Although the scaling law $\gamma=\beta(\delta-1)$ is still preserved, other three scaling laws in Eq.(\ref{scalingl}) are violated.
Obviously, the violation of scaling laws is directly due to $a_{11}=0$, where the underlying physical reason is related to the quantum anomaly effect. Our results illustrate that quantum anomaly can trigger the violation of scaling laws.

Similarly, for our black hole with quantum anomaly in $a_{11}=0$ case, 
we can further rewrite the expression of pressure Eq.(\ref{P}) as
\begin{eqnarray}
\tilde{P}(t,w)&=&\frac{12(t+1)}{5 (w+1)^{1/3}}-\frac{6}{5(w+1)^{2/3}}-\frac{4 (t+1)}{5(w+1)}\nonumber \\&+&\frac{3}{5 (w+1)^{4/3}}.\quad \label{tildepre}
\end{eqnarray}
where $\alpha_{c}=\frac{Q^{2}}{8}$ is used and
the critical values Eq.(\ref{crit}) become
\begin{equation}
    P_{c}=\frac{5}{72\pi Q^{2}},~V_{c}=\sqrt{6}\pi Q^{3},~T_{c}=\frac{1}{\sqrt{6}\pi Q}.
\end{equation}
We expand the dimensionless pressure $\tilde{P}$ in Eq.(\ref{tildepre}) around the critical point like Eq.(\ref{Pexapnsion}), and obtain
\begin{eqnarray}
\tilde{P}(t,w)=1+\frac{8}{5}t-\frac{4}{15}tw^2-\frac{8}{135}w^3+\mathcal{O}(tw^3,w^4),\label{expansion}
\end{eqnarray}
where we find $a_{03}=-\frac{8}{135}$ is negative that means our system is thermodynamically stable. Besides, as shown in the above general discussions, we obtain the four critical exponents of this black hole which are just those results in Eq.(\ref{ce}). Most significantly, we first discover a gravitational system to violate the scaling laws in four dimensional spacetime.

\begin{figure}[h!]
\includegraphics[width=0.45\textwidth]{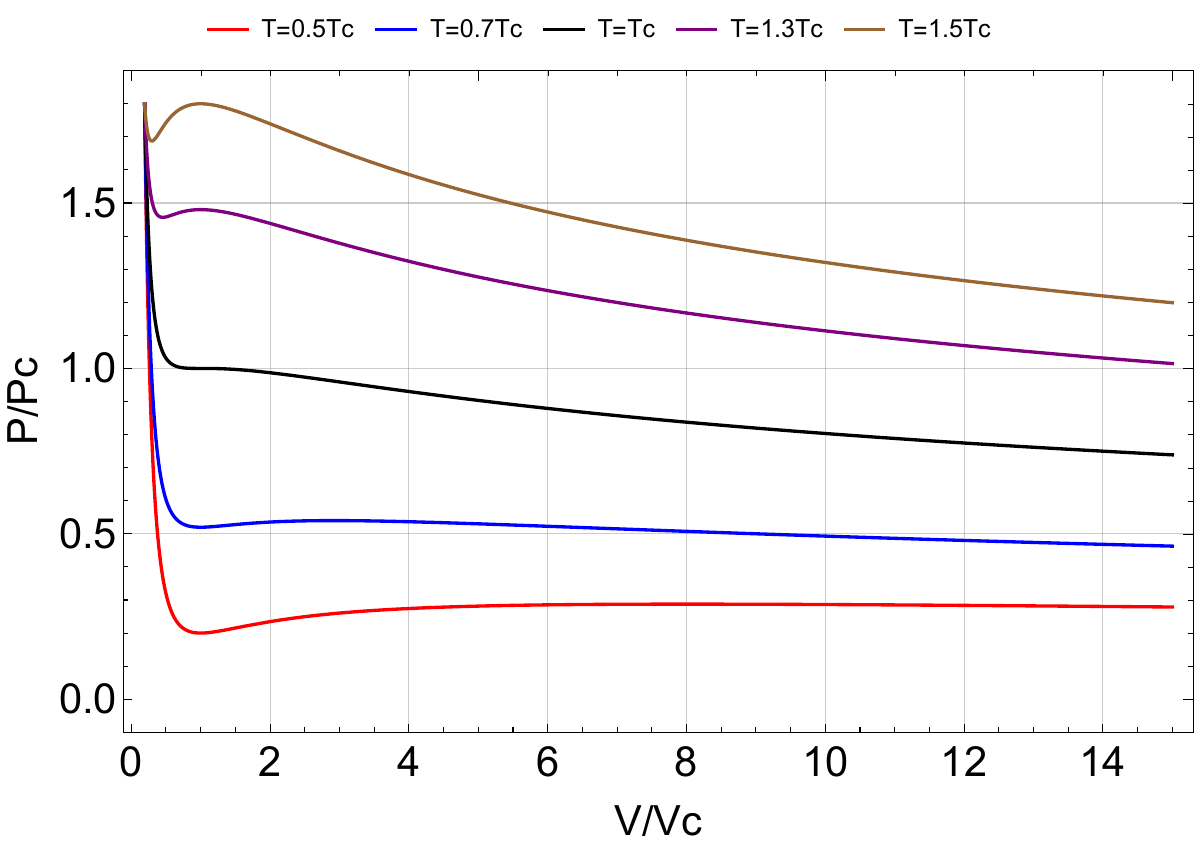}
\caption{Isothermal curves for the black hole with quantum anomaly satisfying condition $a_{11}=0$. The Red/Blue/Black/Purple/Brown lines correspond to $T=0.5T_{c}$, $T=0.7T_{c}$, $T=T_{c}$, $T=1.3P_{c}$ and $T=1.5T_{c}$ respectively, which clearly exhibits a novel property that there are first order phase transitions both below and above the critical temperature. } \label{conformal}
\end{figure}

\begin{figure}[h!]
\includegraphics[width=0.45\textwidth]{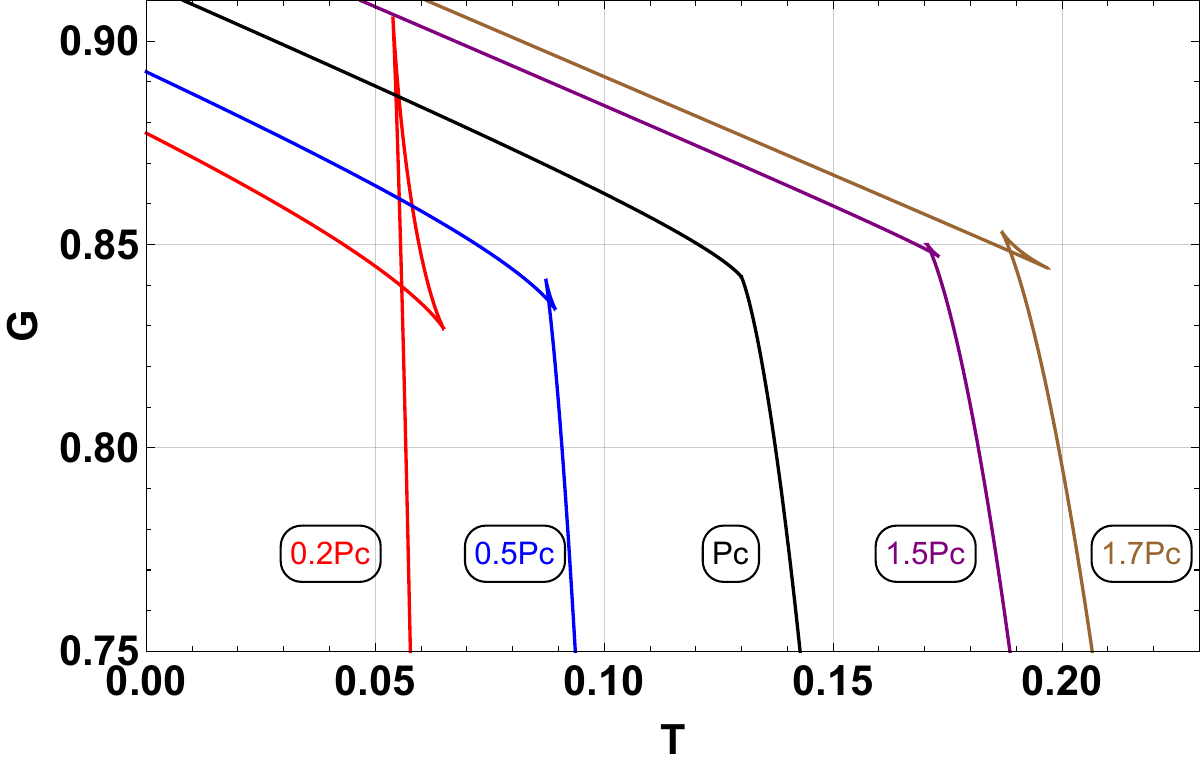}
\caption{Swallowtail behavior of Gibbs free energy at constant pressure.  Red/Blue/Black/Purple/Brown lines  correspond to $P=0.2P_{c}$, $P=0.5P_{c}$, $P=P_{c}$, $P=1.5P_{c}$ and $P=1.7P_{c}$ respectively, and charge $Q$ is setted to $Q=1$.  The Gibbs free energy directly presents the novel phase transition behavior where swallowtail appears both below and above the critical pressure. } \label{conformal1}
\end{figure}

\begin{figure}[h!]
\includegraphics[width=0.45\textwidth]{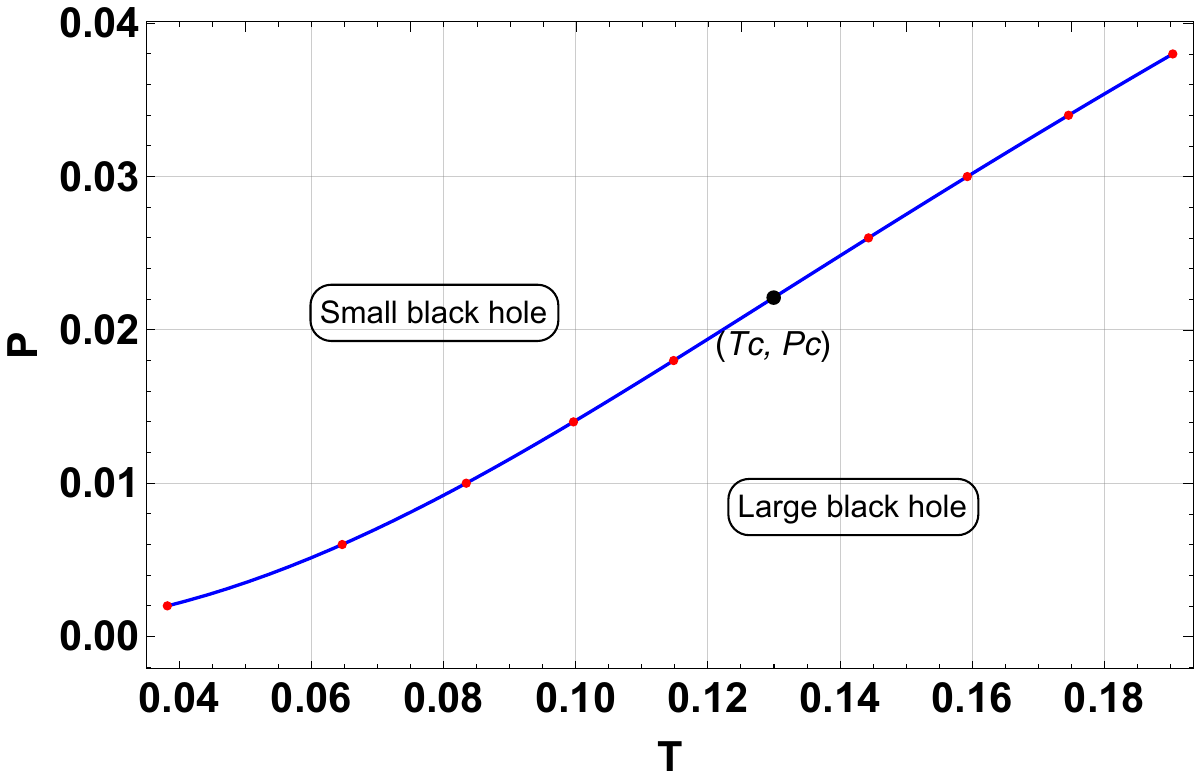}
\caption{Coexistence curve (P-T diagram) between the small and large black hole phases, and black dot denotes the critical point with second order phase transition. Impressively, there are both first order phase transitions below and above the critical point.
} \label{conformal2}
\end{figure}

Besides the violation of scaling laws, we also find a novel property of the phase transition for this black hole with quantum anomaly satisfying $a_{11}=0$. In order to clearly present this novel property, we first plot the isothermal curves in Fig.\ref{conformal} according to the dimensionless equation of state Eq.(\ref{tildepre}), where we can easily find the novel property that the first order phase transition appears both below and above the critical temperature.
The swallowtail behavior of Gibbs free energy is also helpful to show this novel property, and the precise Gibb free energy in $a_{11}=0$ case is
\begin{eqnarray}
    G&=&\frac{1}{6r_{h}}[-6\alpha_{c}+r_{h}^{2}(8\pi P r_{h}^{2}-6\pi T r_{h}+3)\nonumber\\
    &+&48\pi \alpha_{c} T r_{h} \ln(\frac{r_{h}}{\sqrt{2\alpha_{c}}})+3Q^{2}].\label{gibbsf}
\end{eqnarray}
We plot the behavior of Gibbs free energy with different constant pressures in Fig.\ref{conformal1}, and indeed clearly find the swallowtail behavior appearing both below and above the critical pressure.
In fact, we can directly see the novelty from the $P$-$T$ phase diagram, which has also been presented in Fig.\ref{conformal2}. 
Impressively, the novelty manifests a very different property from that in Van der Waals like systems such as the above black hole system with nonzero coefficient $a_{11}$ \cite{Wei:2020poh} and RN-AdS black hole \cite{Kubiznak:2012wp}. That is, the coexistence curve still extends above the critical point, while it terminates at the critical point in Van der Waals like systems. This phenomenon is due to the fact that for $\alpha_{c}<Q^{2}/8$ there exist two critical points as shown above, while for $\alpha_{c}=Q^{2}/8$ the two critical points merge into each other. 

Note that, although the similar thermodynamic
critical phenomena were previously explored in seven-dimensional hyperbolic black holes
with K=3 Lovelock terms \cite{Tavakoli:2022kmo,Frassino:2014pha,Dolan:2014vba}, our work establishes three groundbreaking advancements that
fundamentally distinguish it from prior studies. 
The first one is that we break through the limitation of previous works to physical spacetimes which possess spherical horizon (k=+1) and observable universe’s dimensionality (d=4). Furthermore, we provide the first conclusive evidence that quantum conformal anomaly intrinsically modifies scaling relations. This establishes a clear causal link between quantum effects and scaling law violation – a connection absent in all previous Lovelock studies \cite{Tavakoli:2022kmo,Frassino:2014pha,Dolan:2014vba}. Finally, we also discovered the key condition $a_{11}=0$, which paves the way towards finding more systems with the violation of scaling law. These innovations – bridging critical phenomena to observable spacetime dimensions and exposing quantum signatures in scaling law violation –
constitute our work’s definitive contribution to the field. 

{\bf Conclusion and Outlooks}---
\label{dis}
Scaling laws are universal 
in different critical phenomena, which various distinct physical systems obey same relations among critical exponents. 
In this letter,  
however, we discover a general condition for gravitational system to violate the scaling laws, which is the coefficient $a_{11}$ in the expansion of pressure $P(T,V)$ near the critical point satisfying $a_{11}=0$.  
Particularly, we find that $a_{11}=0$ can appear in the black hole system with quantum conformal anomaly, and hence this gravitational system has the ability to violate scaling laws. The physical interpretation of $a_{11}=0$ becomes explicitly evident when one examines the Gibbs free energy landscape for the phase transition process, which makes free energy go beyond the usual mean field theory behavior. We leave the computational details about this point to the Supplemental Materials. Note that this violation of scaling laws can only occur when central charge parameter $\alpha_{c}$ and $U(1)$ charge $Q$ has following relation $\alpha_{c}=\frac{Q^{2}}{8}$ which is directly deduced from the condition $a_{11}=0$, while for $a_{11}\neq 0$ case, this black hole system shows mean field theory behavior. Therefore, these two interesting different behaviors imply the quantum conformal anomaly just triggers the breakdown of scaling laws. Furthermore, in $a_{11}=0$ case, we also found that the black hole system with conformal anomaly clearly presents a novel phase structure. This phase structure is vastly different from that of van de Waals like system. These results open novel avenues for revealing the properties of systems with violation of scaling laws. 

It is worthwhile to give some outlooks on our results. Although we have found a black hole system violating scaling laws, whether there are other four dimensional gravitational systems sharing the same behaviors remain an open issue.
In addition, it is important to further understand the microscopic mechanism for the violation of scaling law. One potential direction towards this question is that we can understand more about the topological viewpoint of black hole phase transition and the vortex anti-vortex annihilation phenomenon as shown in Refs.\cite{Wei:2022dzw,Ahmed:2022kyv}.
Particularly, whether the novel phase transition behavior predicted in this letter can be realized in laboratory is an interesting direction to be pursued. Furthermore, besides our case in this letter, there may be other cases with scaling-law violation such as nonzero $a_{21}$ with $a_{11}=a_{12}=0$, which may present different critical exponents from our results (\ref{ce}). Although this condition will not be satisfied  due to limited parameter space in this letter, it remains an interesting question to seek a gravitational system which can realize this. Progress in those directions will shed significant light on the understanding of gravitational system and renormalization group theory.


\subsection*{Acknowledgement}

Authors thank Profs. Bum-Hoon Lee, Jinwu Ye, Wei-Qiang Chen, Hongbao Zhang, Shao-Wen Wei and Dr. Shi-Bei Kong for their valuable discussions. This work is supported by National Natural Science Foundation of China (NSFC) under Grant Nos. 12175105, 12405066, 12275106, 12235019. Yu-Sen An is also supported by the Natural Science Foundation of Jiangsu Province under Grant No. BK20241376. 

\newpage
\onecolumngrid
\appendix 
\clearpage
\renewcommand{\thefigure}{S\arabic{figure}}    
\setcounter{figure}{0} \renewcommand{\theequation}{S\arabic{equation}}
\setcounter{equation}{0}
\setcounter{subsection}{0}
\renewcommand{\thesubsection}{\Alph{subsection}}
\section*{Supplemental Information}
\subsection*{Physical meaning of condition $a_{11}=0$ viewed from Gibbs free energy landscape}\label{s1}
In this supplemental material, we give a physical interpretation of condition $a_{11}=0$ by examining the Gibbs free energy for the phase transition process. 
We find that the condition $a_{11}=0$ makes free energy go beyond the usual Landau mean field theory result, which can be seen in the followings. 

The Gibbs free energy is computed as Eq.(\ref{gibbsf}) which reads
\begin{equation}
    G=\frac{1}{6r_{h}}[-6\alpha_{c}+r_{h}^{2}(8\pi P r_{h}^{2}-6\pi T r_{h}+3)+48\pi \alpha_{c} T r_{h} \ln(\frac{r_{h}}{\sqrt{2\alpha_{c}}})+3Q^{2}].\label{gibbsfree}
\end{equation}
Note that Eq.(\ref{gibbsfree}) can be further considered as an off-shell Gibbs free energy as the function of $r_{h}$ for fixed $P,T$.  
 The ensemble temperature $T$ is not necessarily the Hawking temperature as we consider the fluctuation of $r_{h}$. 
 The equilibrium state with Hawking temperature appears at the extremal point of $G$ under the condition 
\begin{equation}
(\frac{\partial G}{\partial r_{h}} )_{P,T}=\frac{1}{2}-\frac{Q^{2}}{2r_{h}^{2}}+4\pi P r_{h}^{2}-2\pi r_{h}T+\frac{\alpha_{c}}{r_{h}^{2}}+\frac{8\pi T \alpha_{c}}{r_{h}}=0, 
\end{equation}
which is equivalent to the equation of state, and hence consistent with physical expectation. 

We can choose the dimensionless volume $\omega=V/V_{c}-1$ (which is equivalent to density difference) as the order parameter analogous to the treatment of van der Waals liquid-gas system. Expanding the off-shell free energy in terms of $\omega$ at leading order around critical point gives the following expansion
\begin{equation}\label{free}
    G(P,T,\omega)=G_{0}(P,T)+(\tilde{P}-1-a_{10}t)\omega-\frac{a_{11}}{2}t\omega^{2}-\frac{a_{03}}{4}\omega^{4}+O(t\omega^{3})
\end{equation}
where $\tilde{P}=P/P_{c}$, $t=T/T_{c}-1$ and the coefficients are
\begin{equation}
    a_{10}=\frac{4(-4\alpha_{c}+K+3Q^{2})}{-22\alpha_{c}+9Q^{2}},\quad a_{11}=\frac{-48\alpha_{c}-10 K+6Q^{2}}{-66\alpha_{c}+27Q^{2}}, \quad a_{03}=\frac{40\alpha_{c}+K-15Q^{2}}{27(-22\alpha_{c}+9Q^{2})}
\end{equation}
with $K\equiv\sqrt{192\alpha_{c}^2+9Q^4-96\alpha_{c} Q^2}$. 

\begin{figure}[h!]
\centering
\includegraphics[width=0.7\textwidth]{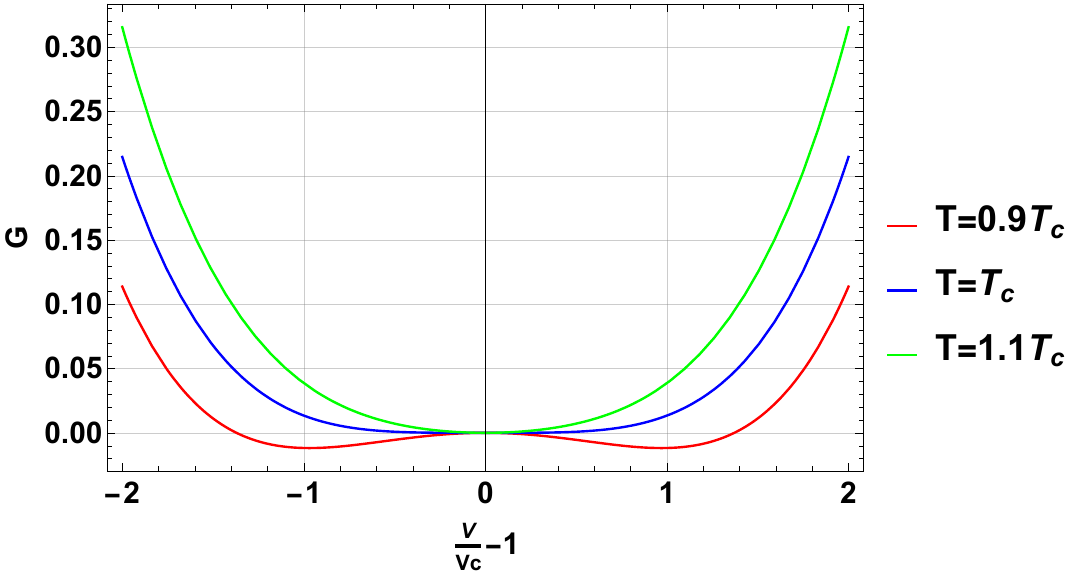}
\caption{Gibbs free energy landscape for $\alpha_{c}=\frac{Q^{2}}{10}$ where $a_{11}\neq 0$. The free energy behavior satisfies Landau mean field theory result where above critical point, two phases merge together. }
\label{gibbs1}
\end{figure}

For $a_{11}\neq 0$ case (where $\alpha_{c}<\frac{Q^{2}}{8}$),when approaching the critical point along the coexistence curve where
\begin{equation}
\tilde{P}=1+a_{10}t+a_{11} t \omega_{l}+a_{03} \omega_{l}^{3}= 1+a_{10}t+a_{11} t \omega_{s}+a_{03} \omega_{s}^{3}=1+a_{10}t, 
\end{equation}
the Gibbs free energy landscape in Eq.(\ref{free}) has the same behavior as Landau mean field theory. That is, at the minimum of free energy, the order parameter vanishes for $T\geqslant T_{c}$ and becomes non-zero for $T<T_{c}$, which have been plotted in Fig.\ref{gibbs1}. In this plot, the temperature is fixed and the pressure is also fixed following the coexistence curve. 

Below the critical temperature, the Gibbs free energy landscape has two minimum with the same free energy which corresponds to the coexistence of small and large black hole. The minimum of free energy is at
\begin{equation}
    \frac{dG}{d\omega}=a_{11}t\omega+a_{03}\omega^{3}=0
\end{equation}
whose roots are given by $\omega_{l,s}=\pm \sqrt{-\frac{a_{11}}{a_{03}}}|t|^{1/2}$ leading to the volume(density) difference between the two phases. Obviously, the two phases merge together at and above the critical temperature. 

\begin{figure}[h!]
\includegraphics[width=0.7\textwidth]{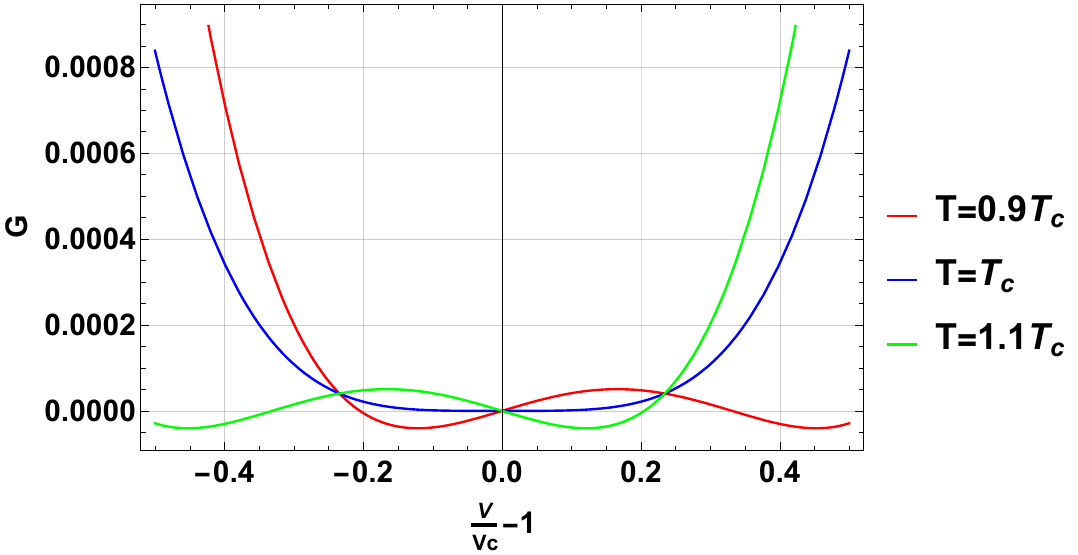}
\caption{Gibbs free energy landscape for $\alpha_{c}=\frac{Q^{2}}{8}$ where $a_{11}=0$. From the free energy behavior, it can be seen that there are both first order phase transition below and above the critical point. }
\label{gibbs2}
\end{figure}

However, for the case   $a_{11}=0$ with $\alpha_{c}=\frac{Q^{2}}{8}$ , in order to further discuss phase transition in this case, we should include higher order term in the free energy which is 
\begin{equation}
    G(P,T,V)=G_{0}(P,T)+(\tilde{P}-1-a_{10}t)\omega-\frac{a_{12}}{3}t\omega^{3}-\frac{a_{03}}{4}\omega^{4}+...
\end{equation}
where coefficients $a_{12}=-\frac{4}{15}$, $a_{03}=-\frac{8}{135}$ and $a_{21}=0$ . This expression is clearly different from the Landau mean field theory like Eq.(\ref{free}). Approaching critical point along the coexistence curve in this case where
\begin{equation}    \tilde{P}=1+a_{10}t+a_{12}t\omega_{l}^{2}+a_{03}\omega_{l}^{3}=1+a_{10}t+a_{12}t\omega_{s}^{2}+a_{03}\omega_{s}^{3}=1+a_{10}t+\frac{2a_{12}^{3}t^{3}}{27a_{03}^{2}},\label{meanfield}
\end{equation}
the term proportional to $\omega$ is non-zero which is also different from $a_{11}\neq 0$ case. 
The behavior of off-shell Gibbs free energy for fixed temperature and pressure in $a_{11}=0$ case is plotted in Fig.\ref{gibbs2} 

The minimum of free energy is
\begin{equation}
    \frac{dG}{d\omega}=\frac{2a_{12}^{3}t^{3}}{27a_{03}^{2}}-a_{12} t \omega^{2}-a_{03}\omega^{3}=0
\end{equation}
whose roots are given by $\omega=\frac{(\sqrt{3}-1)a_{12}t}{3a_{03}}$ and $\omega=-\frac{(\sqrt{3}+1)a_{12}t}{3a_{03}}$. 
The novel property is that, in this case, for both temperatures below and above the critical point, Gibbs free energy bears two minimum with the
same value which corresponds to the coexistence of small and large black holes. This indicates that there are both first order phase transitions below and above critical point which is absent in $a_{11}\neq 0$ case. 

\end{document}